# Symmetry Breaking of Kramers-Henneberger Atoms by Ponderomotive Force

Qi Wei

*State Key Laboratory of Precision Spectroscopy, East China Normal University, Shanghai 200062, China*

Abstract

It was believed that Kramers-Henneberger atoms in superintense laser field exhibit structure of "dichotomy". However this is not true for focused laser field. Because in focused laser, KH state electrons experience ponderomotive force, which will break the dichotomous structure.

Atomic stabilization against ionization is one of the most exotic behaviors for atoms subjected to a superintense high-frequency laser field [1–4]. This phenomenon is termed "adiabatic stabilization" and has fostered many theoretical works [5]. Such atomic stabilization can be understood by means of frame transformation which switches from lab frame to reference frame associated with quiver motion of a classical electron in laser field. The new frame is called "Kramers-Henneberger" (KH) frame, in which electron is at rest and nuclear quivers instead [6-7]. Coulomb potential between electron and nuclear in KH frame is called KH potential, which is dramatically different from original atomic potential. High Frequency Floquet Theory (HFFT) shows that in high-frequency limit, KH potential becomes a cycle-averaged one, termed "dressed" potential, which is time-independent and can support many stable bound eigenstates [1, 3]. Those states are called KH states and the corresponding atoms are called KH atoms. For a KH atom in linearly polarized laser field, dressed potential is similar to a coulomb potential with nuclear charge smeared along a line segment of twice the length $\alpha_0 = \sqrt{I}/\omega^2$, where $I$ and $\omega$ are intensity and frequency of the laser field [1]. Pont *et al.* have shown that at large $\alpha_0$, electron charge density of a ground KH state hydrogen atom is split into two lobes located around the end points of the nuclear charge [8]. This phenomenon has been termed "dichotomy". Structure of "dichotomy" was found not only on single-electron atoms [9-15] but also on two-electron atoms [16-17], not only on ground state but also on low-lying excited states [10-11, 16]. Due to its unique property, dichotomy has even become the symbol of KH atoms and the dichotomous structure has attracted much attention in the past decades [9-21].

However, here we argue that "dichotomy" is not a common structure for KH

atoms in superintense laser fields. Because superintense lasers are practically focused laser beams. In focused laser, the oscillating KH state electron experiences Ponderomotive (PM) force [22-24], which will polarized KH atom and break the symmetry of dichotomous structure.

The time-dependent Schrödinger equation for a one-electron atom in velocity gauge (laboratory frame) is:

$$\left[\frac{1}{2}\left(\boldsymbol{p} - \frac{1}{c}\boldsymbol{A}(t)\right)^2 + V(r)\right]\Psi = i\frac{\partial \Psi}{\partial t} \qquad (1)$$

By applying the KH transformation in the nonrelativistic regime [7, 25], the Hamiltonian can be recast as:

$$\widehat{H} = \frac{\boldsymbol{p}^2}{2} + V(\boldsymbol{r} + \boldsymbol{\alpha}) + \frac{\boldsymbol{A}^2}{2c^2}, \qquad (2)$$

with $\boldsymbol{p}$ the momentum, and $\boldsymbol{\alpha}(t)$ the quiver motion of a classical electron relative to the laboratory frame in laser field. In KH frame, $V(\boldsymbol{r} + \boldsymbol{\alpha})$ is the potential due to interaction of the electron with the nucleus, which corresponds to $V(\boldsymbol{r})$ in the lab frame. $\boldsymbol{A}(t)$ is the vector potential. For an atom subject to a linearly polarized laser pulse propagating along the z direction, the corresponding vector potential has the form:

$$\boldsymbol{A}(t) = \frac{c}{\omega} \cdot E_0(\boldsymbol{r}, t) \cdot \sin[\omega(t - z/c)]\hat{x} \qquad (3)$$

The cycle-averaged vector potential term in the Hamiltonian is actually the PM potential:

$$\widehat{H}_{PM} = \ <\boldsymbol{A}^2/2c^2> \ = |E_0(\boldsymbol{r}, t)|^2/4\omega^2 \qquad (4)$$

which represents the kinetic energy due to oscillation of the electron in the laser field.

If the laser is plane wave, then PM potential is a constant and thus can be abandoned. The other two terms in the Hamiltonian, denoted $\widehat{H}_{KH} = \boldsymbol{p}^2/2 + V(\boldsymbol{r} + \boldsymbol{\alpha})$, defines KH atom, which has been extensively studied in literature [5]. However, KH atom only exists in superintense laser field and superintense laser is practically not plane wave but focused laser beam. Oscillating electrons in focused laser field will experience PM force:

$$F_{PM} = -\nabla\langle\widehat{H}_{PM}\rangle = -\frac{1}{4\omega^2}\nabla|E_0(\boldsymbol{r}, t)|^2 \qquad (5)$$

The PM force comes from non-uniform spatial distribution of PM potential and is only applied on the KH electron. PM force can dramatically change structure of KH atoms and the PM potential term in Hamiltonian should NOT be simply abandoned.

To study the influence of PM force on atomic structure of KH atoms, we take hydrogen atom as an example. In high-frequency limit, the time-dependent dynamics can be converted into a quasistationary Schrödinger equation [1, 3, 7]:

$$\left[\frac{\boldsymbol{p}^2}{2} + V_0(\boldsymbol{r}, \alpha_0) + \boldsymbol{F}_{PM} \cdot \boldsymbol{r}\right]\Phi_{KH} = \epsilon_n(\alpha_0)\Phi \qquad (6)$$

Here the interaction potential has been "dressed" by averaging the electron quiver motion over a laser oscillation cycle:

$$V_0(\mathbf{r}, \alpha_0) = \frac{1}{2\pi} \int_0^{2\pi} \frac{1}{|\mathbf{r} + \boldsymbol{\alpha}(\Theta/\omega)|} d\Theta \qquad (7)$$

with $\boldsymbol{\alpha}(t) = \alpha_0 \cos(\omega t) \hat{x}$. The amplitude, $\alpha_0 = \sqrt{I}/\omega^2$ for linearly polarized laser field, is governed by the laser frequency $\omega$, which is constant, and intensity $I$, which is a function of pulsed laser profile. The validity of Eq. (6) requires that the field must oscillate much faster than electron motion inside the KH atom, which is specified by $\omega \gg \langle \hat{H}_{KH} \rangle$. Since binding energy decreases dramatically with increasing quiver amplitude, Eq. (6) is a suitable approximation for UV or visible light given that quiver amplitude is big enough [14-15]. Eq. (6) can be solved by basis expansion with two-center basis functions:

$$\Phi(\xi, \eta, \phi)_{p,q,m} = (\xi - 1)^p \eta^q [(1 - \eta^2)(\xi^2 - 1)]^{m/2} e^{-\beta \xi} e^{im\phi} \qquad (8)$$

where $p$ and $q$ are non-negative integers and $\beta$ is a variational parameter to optimize the numerical results; $\xi$, $\eta$ and $\phi$ are prolate spheroidal coordinates with $\xi = (r_A + r_B)/2\alpha_0$ and $\eta = (r_A - r_B)/2\alpha_0$.

We assume a linearly polarized laser beam with Gaussian spatial intensity distribution, which are specified in cylindrical coordinates [22, 26]:

$$I(\mathbf{r}) = |E_0(\mathbf{r}, t)|^2 = I_0 \cdot \frac{w_0^2}{w(z)^2} \cdot \exp\left[\frac{-2r^2}{w(z)^2}\right] \qquad (9)$$

where $w(z) = w_0\sqrt{1 + (z/Z_R)^2}$, $w_0$ is the beam waist; $Z_R = \pi w_0^2/\lambda$ is the Rayleigh length; $\lambda = 2\pi c/\omega$ is wave length and ω is laser frequency. Time envelope profile of laser pulse is neglected here.

Take ArF laser ($\omega = 0.24$ a.u.) with peak intensity $I_0 = 1 \times 10^{18}$ W/cm$^2$, as an example. Fig. 1(a-b) shows quiver amplitude $\alpha_0$, radial PM force and eigenenergies of ground KH state hydrogen atom in focal plane (z = 0). PM force increases from zero at focal axis (r = 0) and reaches maximum at half beam size (r = $w_0/2$), then slowly drops to zero when atom moves further away from focal axis. Solid curve in Fig. 1(c) is ground state energy when PM force is not considered. For an atom with radial location within beam waist ($r/w_0 < 1$), binding energy $|\epsilon_0| \leq 0.053$ a.u., which is well below laser frequency $\omega = 0.24$ a.u., so KH state criteria is satisfied. Ground state energy obtained by Eq. (6) is also shown in Fig. 1(c) with PM force along polarization direction (azimuthal angle: $\varphi = 0°$). When $r/w_0 < 1$, it makes a big difference in ground state energy by including the PM force term in Hamiltonian. Fig. 2 shows ground KH state wave functions of hydrogen atom located at focal plane (z = 0) and radial location $r/w_0 = 0.01$, 0.5 and 1.0, respectively. Upper panel shows the famous dichotomy structure when PM force is not considered. Middle panel shows wave functions at laser polarization plane (azimuthal angle: $\varphi = 0°$). In this plane, PM force has maximum projection in laser polarization direction.

The dichotomy structure is polarized by PM force and single-lobe structure is observed. Lower panel shows wave functions in the plane with $\varphi = 80°$. In this plane, the projection of PM force along polarization direction is only $\cos 80° = 0.17$ times of original one. Particularly when $r/w_0 = 0.01$, PM force along polarization direction is as small as $2.7 \times 10^{-6}$ a.u.. Yet the wave function is still polarized by PM force and exhibits single-lobe structure. Results shown in Figs. 1-2 are not unique. Using other lasers than ArF will give similar results.

The easy polarization of KH state hydrogen atom in linearly polarized laser field can be attributed to its unique properties of dressed states. At large quiver amplitude, the dressed potential $V_0$ is the same as that generated by a linear charge. The two lowest-lying states: $\Phi_0$ with symmetry $\sigma_g$ and $\Phi_1$ with symmetry $\sigma_u$, are nearly degenerated [21]. Even a tiny projection of PM force along quiver motion direction will break the symmetry and lift degeneracy. New ground state becomes either $\Phi'_0 = (\Phi_0 + \Phi_1)/\sqrt{2}$ or $\Phi'_1 = (\Phi_0 - \Phi_1)/\sqrt{2}$, depending on the direction of PM force. Both $\Phi'_0$ and $\Phi'_1$ have single-lobe structure located at either end point of the linear charge.

In summary, KH atoms in focused laser beam are polarized by PM force. In linearly polarized laser field and at large quiver amplitude, instead of "dichotomy", KH state H atom exhibits single-lobe structure.

We are grateful for the support from the National Natural Science Foundation of China (Grant No. 11674098).

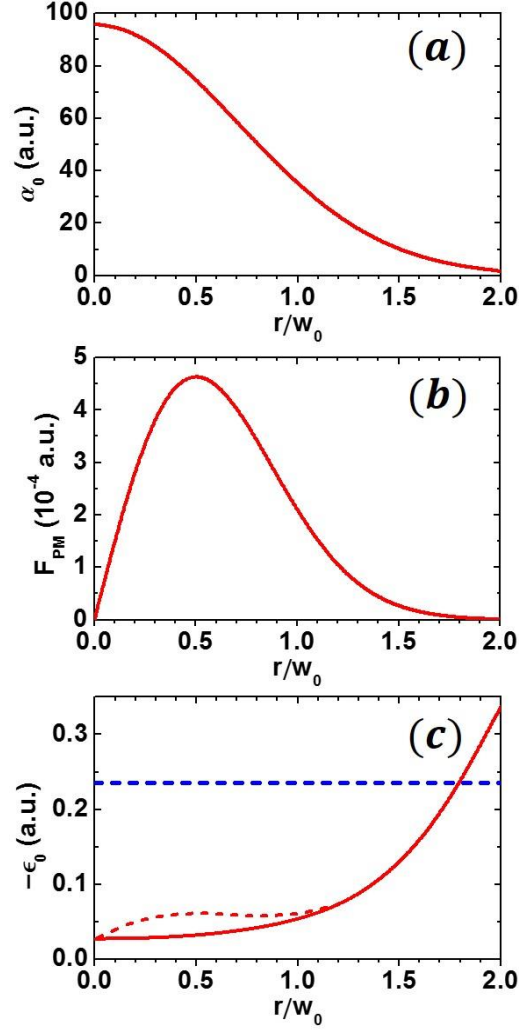

**Fig. 1.** Quiver amplitude (a), radial PM force (b) of ground KH state hydrogen atom in focal plane (z = 0) of pulsed ArF laser as a function of radial position r. The laser has peak intensity $I = 1 \times 10^{18}$ W/cm$^2$ with pulse shape profile of Eq. (9). For panel (c), solid red curve is ground state energy (negative) without considering PM force term; dashed red curve is ground state energy (negative) of Eq. (6) with PM force along polarization direction (azimuthal angle: $\varphi = 0°$); blue dashed line is ArF laser frequency $\omega = 0.24$ a.u..
66

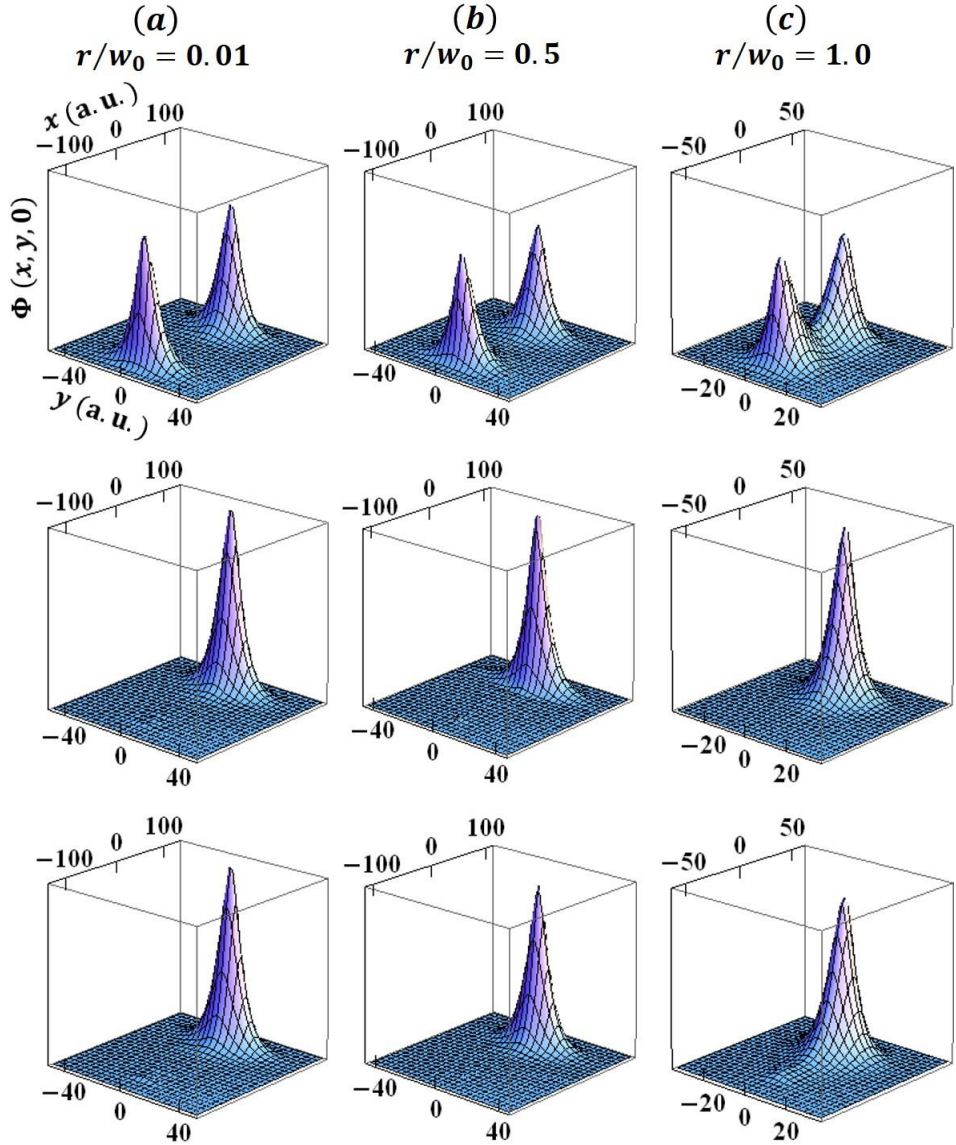

**Fig. 2.** Ground state wave functions of hydrogen atom located at focal plane (z = 0) and radial position $r/w_0 = 0.01$ (panel a), 0.5 (panel b) and 1.0 (panel c), respectively. The laser has peak intensity $I = 1 \times 10^{18}$ W/cm$^2$ with pulse shape profile of Eq. (9). Upper panel shows the famous dichotomy structure when PM force is not considered. Middle panel shows wave functions at laser polarization plane (azimuthal angle: $\varphi = 0°$). Lower panel shows wave functions in the plane with $\varphi = 80°$. Laser is polarized along x direction. For each wave function plot, atom's location is defined as origin ($x = 0, y = 0, z = 0$) and projection of PM force is along $x$ direction.